\documentstyle[12pt]{article}

\begin{document}

\setcounter{page}{1}

\centerline{\large {\bf A Simple Confinement Mechanism for}}
\centerline{\large {\bf Light-Front Quantum Chromodynamics}}
\vskip.3in
\centerline{ Robert J. Perry}
\centerline{ Physics Department, The Ohio State University}
\centerline{ 174 W. 18th Ave., Columbus, Ohio 43210, USA}

\vskip.6 in
\centerline{\bf Abstract}
\vskip.1in

Light-front field theory offers a scenario in which a constituent
picture of hadrons may arise, but only if cutoffs that violate
explicit covariance and gauge invariance are used.  The perturbative
renormalization group can be used to approximate the cutoff QCD
hamiltonian, and even at lowest orders the resultant hamiltonian
displays interesting phenomenological features.  A general scheme for
computing and using these hamiltonians is discussed and it is
explicitly shown that a confining interaction appears when the
hamiltonian is computed to second order.

\vskip.1in

\vskip.3in
\noindent {\bf 1. Introduction and Basic Strategy}
\vskip.1in

Quantum chromodynamics (QCD) is widely accepted as the fundamental
theory of the strong interaction, but we are still unable to solve
this theory in the low energy regime and obtain an accurate
approximation for the structure of hadrons.  This problem remains one
of the most important unsolved problems in physics.  The primary
source of difficulty is that manifestly covariant and gauge invariant
formulations of QCD yield a picture of hadrons as complicated
many-body excitations on top of a complicated vacuum.  In this picture
we must solve coupled strongly-interacting many-body problems to
obtain hadrons, in contrast to the simple few-body states found in the
phenomenologically successful constituent quark model.  Light-front
field theory offers an alternative in which hadrons may be
approximated as few-body bound states.

How can a constituent picture of hadrons arise in light-front field
theory?  The key is renormalization. [1]  The issue is not whether
hadrons contain arbitrarily many quarks and gluons added to a
complicated vacuum.  They do.  The issue is whether the dynamical
effects of almost all of the partons in a hadron and in the vacuum can
be approximated by effective interactions in a light-front QCD
hamiltonian that can be used to compute the dominant `valence'
structure of hadrons.  To address this issue I would like the reader
to consider how many-body states enter the structure of a hadron.

First, in a field theory with local interactions high energy many-body
states do not decouple from low energy few-body states. If one uses
perturbation theory to estimate the errors made when high energy
states are simply removed by a cutoff, the answer is simple. The
errors are infinite.  For example, a single quark mixes
perturbatively with high energy quark-gluon states, and the energy
shift in second-order perturbation theory is infinite.  This is an old
story in field theory.  We know that a regulator must be introduced,
and if these high energy components are to be removed from the state
so that it can be dominated by few-body components, this regulator
must be a cutoff.  However, after a cutoff is introduced, results are
strongly dependent on the cutoff.

To remove cutoff dependence and properly account for the effects of
high free-energy components, renormalization is required. The
`effective' hamiltonian becomes cutoff dependent, and must be designed
to remove cutoff dependence from physical quantities such as masses
and form factors.  If the regulator does not violate the symmetries of
the original theory, the resultant `counterterms' will also respect
these symmetries; and the only relevant and marginal operators that
appear in the effective hamiltonian are canonical masses and
couplings.  Regulators that respect all the symmetries of QCD do not
remove high energy states, and therefore do not yield a constituent
picture.  To obtain a constituent picture we are forced to use cutoffs
that violate these symmetries.  As a result, the effective QCD
hamiltonian will contain operators that also violate these symmetries,
forcing us to invent a renormalization procedure capable of
identifying the extra relevant and marginal operators and fixing their
strength.  Wilson's renormalization group [2,3] suitably generalized
for our problem offers the necessary tools.

Even if we can remove high free-energy components from physical
states, and replace them with effective interactions, we are still
faced with the fact that low free-energy many-body states do not
typically decouple from low free-energy few-body states in QCD. The
vacuum is supposed to be a complicated superposition dominated by low
free-energy states.  This second problem is what forces us to use
light-front coordinates.  In these coordinates we may be able to force
the many-body states that appear as low free-energy states in equal
time field theory to act like high energy states, so that the problem
of replacing them with effective interactions resembles the
renormalization problems we encounter when removing high free-energy
states.

The principal observations that lead to this possibility are simple.
First, the longitudinal momenta conjugate to the light-front
longitudinal spatial coordinate are all positive,

\begin{equation}
p_i^+ \ge 0 \;.
\end{equation}

\noindent  Since every individual longitudinal momentum is positive,
any state with many partons must contain some `wee' partons; {\it
i.e.}, partons with small longitudinal momentum fraction.  The free
energy of a particle in light-front coordinates is

\begin{equation}
p_i^- = {{\bf p}_{\perp i}^2 + m^2 \over p_i^+} \;.
\end{equation}

\noindent This dispersion relation implies that a particle with small
longitudinal momentum is a high energy particle, so that states
containing wee partons are high energy states.  Thus, we are left with
the hope (possibly naive) that if we can successfully remove all high
energy states in QCD and replace them with effective interactions, we
will be left with few-body states and a constituent picture of hadrons
in light-front QCD.

It should be clear that the first step in this program is the
calculation of effective interactions that result from the removal of
high energy partons.  This is the type of problem that led Wilson to
develop his version of the renormalization group.  For details I refer
the reader to the excellent review articles by Wilson and Kogut [2,3],
my recent article on light-front renormalization groups [4], and the
recent articles by G{\l}azek and Wilson [5] that develop the new
similarity renormalization group.  To identify the effective cutoff
hamiltonian, we can directly study the cutoff dependence of the
hamiltonian itself.  The principal tool for this study is a
renormalization group transformation.  Given a hamiltonian with cutoff
$\Lambda_0$, the transformation produces a new hamiltonian with cutoff
$\Lambda_1$. These hamiltonians must produce equivalent results in
some sense, and in the similarity renormalization group they are
unitarily equivalent.  By studying the properties of the
transformation, we can try to identify a cutoff hamiltonian that is
produced by an infinite number of transformations.  If we find such a
hamiltonian, by design it will produce the same results as a
hamiltonian with an infinite cutoff; so it is a renormalized
hamiltonian.

If the cutoff respects all the symmetries of the theory, the cutoff
renormalized hamiltonian should be uniquely identified up to a few
free masses and couplings, and irrelevant operators that can be
ignored if the cutoff is sufficiently large.  In massless QCD, a
single running coupling will remain undetermined.  On the other hand,
if the cutoff violates these symmetries there will be many new
candidate renormalized hamiltonians, because there are many relevant
and marginal operators that violate the symmetries.   Only one of
these hamiltonians should restore these symmetries to physical
quantities; so one strategy for finding the correct hamiltonian is to
identify the relevant and marginal symmetry-breaking operators and
tune their strengths to restore the symmetries.  As far as I know, no
exceptions to these rules have been found.  The basic idea is that the
complete set of symmetries determines the theory.

This procedure is confronted with serious problems in light-front
QCD.  First, in light-front field theory there are an infinite number
of relevant and marginal operators because functions of longitudinal
momenta appear in these operators. [1,4] This problem is due to the
fact that longitudinal scaling is a boost, which is a Lorentz symmetry
that cannot be broken.  While boost invariance should be restored by
only one choice of these functions, apparent ambiguities arise at
finite orders of perturbation theory.  The second problem is unique to
non-abelian gauge theories.  Many gauge-variant Green's functions are
infrared divergent in QCD, and it is difficult to invent a scheme that
can produce all the required counterterms without computing such
Green's functions ({\it e.g.}, the quark-gluon vertex) at an
intermediate stage.  Wilson and I have devised {\it coupling
coherence} to circumvent these problems. [6]

I refer to reader to the literature for details on coupling coherence,
[4,6,7] and will provide only a sketch.  The basic idea is that only
the canonical masses and couplings should be independent functions of
the cutoff, if the cutoff free hamiltonian respects the symmetries of
the theory.  All new relevant, marginal, and irrelevant couplings
should depend on the cutoff only because they depend on these
canonical couplings.  The renormalization group equations determine
how all the constants change with the cutoff; and when one inserts the
ansatz that only a few constants depend on the cutoff, the remaining
constants (including functions of longitudinal momentum fractions) are
determined by the renormalization group equations.  In practice we
have only been able to apply coupling coherence to the perturbative
renormalization group; but in all cases considered to date, the
counterterms that result are exactly those required to restore the
symmetries of the theory, even though no direct reference is made to
these symmetries in the calculations. [4,6]

The conclusion we have reached is that, given a cutoff, the
renormalization group and coupling coherence uniquely determine the
hamiltonian to each order in the canonical coupling.  In QCD this
allows us to compute the effective hamiltonian as an expansion,

\begin{equation}
H_\Lambda = H^{(0)} + g_\Lambda H^{(1)} + g^2_\Lambda H^{(2)} +
\;\cdot \cdot \cdot \;.
\end{equation}

\noindent I have suppressed the fact that there is also dependence on
the running current quark masses; but the most important point is that
the operators, $H^{(n)}$, depend on $\Lambda$ only because of their
dimension ({\it e.g.}, a factor of $\Lambda^2$ for mass operators) and
because of the cutoff functions.  If the cutoff is chosen properly, as
discussed below, we may be able to exploit asymptotic freedom to
approximate the QCD hamiltonian by truncating this series at a finite
order.  It will almost certainly be necessary to further tune the
strength of the relevant operators ({\i.e.}, the quark and gluon
dispersion relations, and the chiral symmetry breaking quark-gluon
vertex); but this is our starting point.

Having computed an approximate cutoff QCD hamiltonian, the next step
is to study this hamiltonian non-perturbatively.  An essential part of
this step is the demonstration that the resultant low energy states
are indeed dominated by few-body components so that a constituent
picture arises.  I am going to oversimplify this second step by
ignoring the fact that we will need to push the cutoff as low as
possible and confront the fact that the coupling becomes large as
$\Lambda \rightarrow \Lambda_{QCD}$.  Elsewhere in these proceedings,
Wilson and Robertson discuss a strategy in which a sequence of
weak-coupling calculations are extrapolated to this potentially large
coupling, [8] a strategy first outlined by Wilson and collaborators.
[1] For the purpose of this article it is sufficient to assume that
the cutoff remains sufficiently large that the coupling does not
become unmanageably large.  If this is the case, we can simply use
bound state perturbation theory to study our approximate QCD
hamiltonian.

Once the cutoff is lowered to a suitable point where it is conceivable
that the important many-parton components of hadron wave functions
have been `integrated out,' we must still deal with remaining
interactions that involve parton emission and absorption.  We assume
that these interactions become unimportant at small cutoffs, so that
it is possible to first approximate the hamiltonian by keeping only
interactions in which parton number is conserved.  This ansatz is
quite reasonable from a variational point of view.  If we consider a
trial state in which a quark-antiquark pair are separated, the
expectation value of the hamiltonian provides an upper bound on the
energy.  Any additional quarks and gluons in the wave function can
only lower the energy.  This means that if the hamiltonian is
confining, there must be a two-body interaction which causes the
energy to grow without bound as the pair is separated.  In other
words, the type of interactions we need to get a reasonable
phenomenology without parton emission and absorption must actually
appear in the hamiltonian, although there is no guarantee that they
will appear in a perturbative approximation to the hamiltonian.  We
assume that few-body interactions largely determine the structure of
hadrons, so that the additional interactions can be studied in bound
state perturbation theory.

Given any hamiltonian, one can study bound states by first writing

\begin{equation}
H=H_0+V \;.
\end{equation}

\noindent To paraphrase Weinberg, you are free to choose any $H_0$ you
please; but if you choose wrong, you'll be sorry.  The main criteria
in choosing $H_0$ are first that it be a reasonable approximation of
$H$, so that bound state perturbation theory does not diverge; and
second that it be solvable.  Since the problem we will initially
address is that of meson structure, this last restriction simply means
that the quark-antiquark and quark-antiquark-gluon problems with
two-body interactions should be tractable.  This is not an overly
severe restriction if one is willing to use a computer.

The strategy I will follow in this paper mirrors the strategy outlined
above.  I will compute the QCD hamiltonian to ${\cal O}(g^2)$ using
a similarity renormalization group and coupling coherence, and I will
then show that with a reasonable choice of $H_0$ this hamiltonian
confines quarks and gluons. This result was first shown in Ref. [7].

There are two related questions one must ask to decide if the simple
confinement mechanism survives.  We first compute $H$ perturbatively
by removing high energy states, and we must ask whether confining
interactions that appear at low orders in this calculation survive to
higher orders.  We then use bound state perturbation theory to study
the approximate hamiltonian, and this depends on an explicit choice of
$H_0$.  We must ask whether a choice of $H_0$ which includes confining
interactions from $H$ leads to a reasonable bound state perturbative
expansion.  These questions are actually intertwined, but it is easier
to study the second question using order-of-magnitude arguments than
it is to study the first.

I must emphasize that to second order one can also force QED to be
confining; but it is relatively straightforward to see that if $H_0$
is chosen to contain the confining interactions in QED, there are
large perturbative corrections that cancel confinement.  The
fundamental observation is that even when confinement is included in
$H_0$ for QED, photons are massless and not confined.  This means
photon exchange persists to arbitrarily large distances as charged
particles separate and this photon exchange cancels the confining
interaction.  On the other hand, in QCD the same interactions that
confine quarks appear in second order to confine gluons.  This means
that it is self-consistent to assume that the confinement mechanism
survives because confinement turns off the long-range gluon exchanges
that are necessary to cancel confinement.  Hopefully this point will
be clarified somewhat below.

\vskip.3in
\noindent
{\bf 2. Quark and Gluon Dispersion Relations from Coupling Coherence}
\vskip.1in

The problem I want to address in this Section is the calculation of
the one-body operators ({\i.e.}, the quark and gluon dispersion
relations) to second-order in the QCD coupling constant.  Since this
constant does not run until third order, I only need to consider how
the coupling runs to justify my choice of cutoffs.  I can use any
cutoff I want, but I would like to have some hope that the
second-order results are not meaningless.  This means I have to
exploit asymptotic freedom to justify the first step in the analysis,
the perturbative calculation of $H$.

If the cutoff is chosen properly, the QCD hamiltonian is approximately
free when the cutoff is large.  The free hamiltonian must be a fixed
point ({\it i.e.}, a hamiltonian that does not change under the action
of the transformation) for this to happen, which is actually rather
easy to arrange since the transformation reduces to a scaling
operation when applied to free hamiltonians.  Near this fixed point,
degrees of freedom with nearly the same free energy may still couple
strongly to one another even when the coupling constant is small,
which follows from nearly degenerate perturbation theory.  Degrees of
freedom that have drastically different free energy couple weakly.
This means that if we want to exploit the fact that the coupling
constant is small, the cutoff cannot remove the coupling between
nearly degenerate degrees of freedom.

If the cutoff cuts through nearly degenerate degrees of freedom, we
must solve a non-perturbative problem to replace the effects of their
coupling with effective interactions.  This is exactly what I want to
avoid.  Therefore, I am forced to use a cutoff on free energies.  If
the cutoff removes states ({e.g.}, all particles with a free energy
above some fixed value are removed), states just below the cutoff will
couple strongly to states just above the cutoff, and we must again
solve a non-perturbative problem to replace the effects of their
coupling with effective interactions.  For example, the quarks in a
high energy pion interact strongly with one another.  Therefore I am
forced to use a cutoff that does not remove states, but instead
removes only the direct coupling between states of drastically
different free energy.  In other words, the cutoffs must act at the
vertices, preventing the free energy from changing by more than a
fixed amount through a vertex.  This is exactly the type of cutoff
that the similarity transformation runs.  In summary, to exploit
asymptotic freedom I must use a cutoff that removes the coupling
between states of drastically different free energy.  The easiest
cutoff functions to use in low order analytic calculations are step
functions, which I will use here; although step functions introduce
pathologies that are undesirable later.

Before proceeding to a brief discussion of the similarity
transformation and coupling coherence, I want to point out a very
interesting feature of the cutoff on free light-front energies.
Light-front energy has the dimension of transverse momentum squared
(the same as mass squared) divided by longitudinal momentum.
Generically, our cutoff is $\Lambda^2/{\cal P}^+$, where $\Lambda$ has
the dimension of mass.  Transverse scale invariance is violated,
leading to dimensional transmutation, and the QCD coupling constant is
forced to run with $\Lambda$.  However, our cutoff is $\Lambda^2/{\cal
P}^+$ and contains an arbitrary longitudinal momentum scale, ${\cal
P}^+$.  If we succeed in renormalizing the theory, $\Lambda$
dependence will disappear from physical quantities, which means that
${\cal P}^+$ dependence will also disappear; but this will not happen
exactly in a perturbative approximation, and because of this we will
also find dependence on ${\cal P}^+$ in perturbative approximations.
The appearance of this extra longitudinal momentum scale has profound
implications for our program, some of which are illustrated and
discussed below.

In order to use coupling coherence I need to study how the hamiltonian
changes when the cutoff changes.  I want to avoid a detailed
derivation of a similarity transformation, [5] so I will just give the
result I need through second order and show that it is easily
understood.  Let $H=h_0+v$, where $h_0$ is a free hamiltonian and $v$
is cut off so that

\begin{equation}
\langle \phi_i | v | \phi_j \rangle = 0 \;,
\end{equation}

\noindent if $|E_{0i}-E_{0j}| > \Lambda$; where $h_0 |\phi_i\rangle =
E_{0i} |\phi_i \rangle$.  If this cutoff is lowered to $\Lambda'$, the
new hamiltonian matrix elements to ${\cal O}(v^2)$ are

\begin{eqnarray}
H'_{ab} &=&
\langle \phi_a|h_0+v|\phi_b\rangle \nonumber \\
&-& \sum_k v_{ak} v_{kb} \Biggl[
{\theta\bigl(|\Delta_{ak}|-\Lambda' \bigr)
\theta\bigl(|\Delta_{ak}|-|\Delta_{bk}|\bigr) \over E_{0k}-E_{0a} }
+ {\theta\bigl(|\Delta_{bk}|-\Lambda'\bigr)
\theta\bigl(|\Delta_{bk}|-|\Delta_{ak}|\bigr) \over E_{0k}-E_{0b} }
\Biggr] ,
\end{eqnarray}

\noindent where $\Delta_{ij}=E_{0i}-E_{0j}$ and $|E_{0a}-E_{0b}|<
\Lambda'$.  To follow the details of the discussion it is important to
remember that there are implicit cutoffs in this expression because
the matrix elements of $v$ have already been cut off so that
$v_{ij}=0$ if $|E_{0i}-E_{0j}|>\Lambda$. There are actually an
infinite number of similarity transformations that will reduce the
cutoff on how far off the diagonal matrix elements appear, but I will
not discuss the additional constraints I have placed on the
transformation to arrive at this result.  They are not central to my
discussion.  I should note that I have fixed an error in my Brasil
lectures, [7] where I used a transformation that does not completely
avoid small energy denominators.

It is rather easy to understand this result qualitatively.  We have
removed the coupling between degrees of freedom whose free energy
difference is between $\Lambda'$ and $\Lambda$, so the effects of
these couplings are forced to appear in the new hamiltonian as direct
interactions.  To first order, the new hamiltonian is the same as the
old hamiltonian, except that couplings between $\Lambda'$ and
$\Lambda$ are now zero.  To second order, the new hamiltonian
contains a new interaction which sums over the second-order effects of
couplings that have been removed.  The second-order term in the new
hamiltonian resembles the expression found in second-order
perturbation theory, which is not surprising since the new hamiltonian
must produce the same perturbative expansion for eigenvalues, cross
sections, etc. as the original hamiltonian.

I have chosen the transformation so that it is always possible to find
a coupling coherent hamiltonian to second order.  To this order, we
want the hamiltonian to reproduce itself, with the only change being
$\Lambda \rightarrow \Lambda'$.  The solution is found by noting that
we need the partial sum above to be added to an interaction in $v$
that is expressed as a sum, so that the transformation merely changes
the limits on the sum in a simple fashion.  There are two
possibilities.  The first is

\begin{eqnarray}
H_{ab} &=&
\langle \phi_a|h_0+v|\phi_b\rangle \nonumber \\
&-& \sum_k v_{ak} v_{kb} \Biggl[
{\theta\bigl(|\Delta_{ak}|-\Lambda \bigr)
\theta\bigl(|\Delta_{ak}|-|\Delta_{bk}|\bigr) \over E_{0k}-E_{0a} }
+ {\theta\bigl(|\Delta_{bk}|-\Lambda\bigr)
\theta\bigl(|\Delta_{bk}|-|\Delta_{ak}|\bigr) \over E_{0k}-E_{0b} }
\Biggr] ,
\end{eqnarray}

\noindent and the second is

\begin{eqnarray}
H_{ab} &=&
\langle \phi_a|h_0+v|\phi_b\rangle \nonumber \\
&+& \sum_k v_{ak} v_{kb} \Biggl[
{\theta\bigl(\Lambda-|\Delta_{ak}| \bigr)
\theta\bigl(|\Delta_{ak}|-|\Delta_{bk}|\bigr) \over E_{0k}-E_{0a} }
+ {\theta\bigl(\Lambda-|\Delta_{bk}| \bigr)
\theta\bigl(|\Delta_{bk}|-|\Delta_{ak}|\bigr) \over E_{0k}-E_{0b} }
\Biggr] .
\end{eqnarray}

\noindent Note that the $v$ in these expressions is the same as that
above only to first order.  The coupling coherent interaction in $H$
is written as a power series in $v$ which reproduces itself under the
transformation, except the cutoff changes.  In higher orders the
canonical variables would also run.

Given the generic coupling coherent hamiltonian to second order, it is
a conceptually simple exercise to compute the coherent QCD hamiltonian
to second order.  For a second-order calculation it is sufficient to
assume that $v$ contains all canonical QCD interactions. Space does
not permit me to list the canonical QCD hamiltonian, so I must again
refer the reader to the literature for details. [1,7,9,10] It is not
necessary to be careful in the derivation of the canonical
hamiltonian, because coupling coherence will take care of details.  It
is sufficient to naively derive the canonical hamiltonian in
light-cone gauge, $A^+=0$, and insert cutoffs on free energy transfer
in each of the vertices.  The next step is to compute the ${\cal
O}(g^2)$ corrections using Eq. (7) or Eq. (8).  To decide which of
these equations to use one must in principle go to higher orders, but
in practice it is usually obvious which choice is correct.  In the
remainder of this section I will discuss the ${\cal O}(g^2)$
corrections to the one-body operators in the QCD hamiltonian.

First consider the second-order correction to the quark self-energy.
This results from the quark mixing with quark-gluon states whose
energy is above the cutoff.  If we assume that the light-front energy
transfer through the quark-gluon vertex must be less than
$\Lambda^2/{\cal P}^+$, the coupling coherent self-energy for quarks
with zero current mass is

\begin{eqnarray}
\Sigma_{\Lambda}(p)&=& {g_\Lambda^2 C_F \Lambda^2 \over 4 \pi^2 {\cal
P}^+} \Biggl\{ \ln\Biggl({ p^+ \over \epsilon {\cal P}^+} \Biggr) - {3
\over 4} \Biggr\} + {\cal O}(\epsilon) \;.
\end{eqnarray}

\noindent Let me first describe the variables that appear in this
result and then turn to a discussion of two important features. The
quark has longitudinal momentum $p^+$, while the longitudinal momentum
scale in the cutoff is ${\cal P}^+$.  This coupling coherent solution
comes from the second generic solution above, Eq. (8), in which one
sums over states below the cutoff.  This sum becomes an integral in
the continuum theory, and I have completed the quark-gluon loop
integral to obtain this result.

The first and most interesting feature of this result is that I have
been forced to introduce a second cutoff,

\begin{equation}
p_i^+ > \epsilon {\cal P}^+ \;,
\end{equation}

\noindent which restricts how small the longitudinal momenta of any
particle can become.  Without this second cutoff on the loop momenta,
the self-energy is infinite, even with a cutoff on free energies.
This second cutoff should be thought of as a longitudinal resolution.
As we let $\epsilon \rightarrow 0$ we resolve more and more wee
partons, and in the process we should confront effects normally
ascribed to the vacuum.  In this case the wee gluons are responsible
for giving the quark a mass that is literally infinite.  Theorists who
insist on deriving intuition from manifestly gauge invariant
calculations may find this interpretation repugnant, but within the
framework of a light-front hamiltonian calculation it is quite
natural.  It is gauge invariance that is not natural, a heretical
conclusion that will put light-front theorists on the defensive until
we solve non-perturbative problems that have not been solved with
other methods.

This second, infrared cutoff poses a problem.  If we introduce a
second cutoff, shouldn't we introduce a second renormalization group
transformation to run this cutoff and find the new counterterms
required by it?  The oversimple answer I will need here is `no.'  The
divergences that require us to introduce $\epsilon$ appear only in
second-order diagrams and subdiagrams, so they look like
super-renormalizable divergences that can be removed to all orders by
a few counterterms.  In principle the infrared divergences could
require us to introduce complicated functions of transverse momentum,
a possibility emphasized by Wilson.  However, in perturbation theory
we find that these divergences always cancel without the need for
counterterms that violate transverse locality. [11] I will assume here
that we can maintain such cancellations at all stages of our
calculation.

When we compute $H_\Lambda$ perturbatively, the cancellations are
those of perturbation theory.  For example, the divergence in the
quark mass is canceled by perturbative mixing of the quark with
quark-gluon states, until the cutoff approaches $\Lambda_{QCD}$.
There is no phenomenological reason to believe that such a
cancellation can persist as the cutoff approaches $\Lambda_{QCD}$,
because there are no free massless gluons.  This means that the
perturbative cancellations at high energies must be replaced by new
cancellations at low energies, cancellations that do not require
mixing between few-body and many-body states. [11]  These
cancellations are related to confinement, as we will see.  When we
study the cutoff hamiltonian in bound state perturbation theory the
need to maintain precise cancellation of all infrared divergences
places severe constraints on our choice of $H_0$.

So, for the purposes of this paper, I will assume that the infrared
divergences are simple enough that we can introduce the cutoff
$\epsilon$, and take it to zero at the end of the calculation.  The
reason that this answer is oversimple is because eventually one
discovers that parton-parton interactions diverge as longitudinal
momenta go to zero.  This can be understood by thinking about the fact
that the cutoff is $\Lambda^2/{\cal P}^+$, so that reducing the
longitudinal momentum of a pair of partons that interact is equivalent
to lowering the cutoff $\Lambda$; and $g_\Lambda$ increases as
$\Lambda$ decreases.  This issue is extremely important, and I am
avoiding it because it is complicated and because I do not have a full
solution to this problem.  I will only add one cryptic remark.  If a
renormalization group is used to run a cutoff on longitudinal momenta,
the full interacting QCD hamiltonian must be a fixed point of the
longitudinal transformation because longitudinal scaling is a Lorentz
symmetry that cannot be violated. [4]

If we want to let $\epsilon \rightarrow 0$, we must face the fact that
the quark self-energy diverges even when $\Lambda$ is finite and
identify a new cancellation mechanism.  There are two possibilities.
First, the divergences could be canceled in the energy of a physical
quark.  Second, the energy of a single quark could remain infinite
with the divergences only being canceled in color singlet states.
The first possibility is clearly the one required in QED.  However,
there is no experimental evidence for a finite mass quark; so we can
explore the possibility that only color singlets have finite mass.
This can only happen if there are infrared divergent interactions that
exactly cancel the infrared divergent part of the self-energy, and I
will show that this does indeed happen in a second-order analysis of
QCD.  Roughly speaking, the self-energy of a monopole diverges but
that of a neutral dipole does not.

The second interesting feature of the above `mass' is that it produces
a dispersion relation which differs from that of current masses.
Normally a mass produces an energy of the form $m^2/p^+$, where $p^+$
is the longitudinal momentum of the parton; but here we find
$\Lambda^2/{\cal P}^+$.  This means that the energy does not diverge
like $1/p^+$, but at this order is independent of the parton
momentum.

A nearly identical calculation reveals the second-order self-energy of
gluons, and again we find that the dominant term goes like

\begin{equation}
{g_\Lambda^2  \Lambda^2 \over {\cal P}^+}
\ln\Biggl({ p^+ \over \epsilon {\cal P}^+} \Biggr)  \;.
\end{equation}

\noindent I do not list the exact expression because it is not
important.  In QED we find that the photon mass in the cutoff
hamiltonian is infrared finite, and we expect that it is exactly
canceled by mixing with electron-positron pairs.  However, in QCD the
gluon mass is infinite and cannot be canceled by mixing with gluon
pairs if there are no free massless gluons.  Here the story is almost
identical to that for quarks.  If this divergence is not canceled by
such mixing, there must be a divergent interaction involving gluons
that allows it to be canceled in color singlet states.  Once again,
this is exactly what happens in a second-order analysis of QCD.

\vskip.3in
\noindent {\bf 3. Confinement from Coupling Coherence}
\vskip.1in

In addition to one-body operators we find quark-quark, quark-gluon,
and gluon-gluon interactions in the second-order coupling coherent
hamiltonian.  As we lower the cutoff, we remove gluon exchange
interactions, and these are replaced by direct interactions.  The
analysis of all of these interactions is nearly identical, so I will
only consider the quark-antiquark interaction.

As stated above, in light-front coordinates partons with small
longitudinal momentum are high energy partons, and this has important
consequences for the light-front hamiltonian.  We find important tree
level counterterms that are not encountered in equal time
coordinates.  In equal time coordinates a high free-energy gluon has a
large momentum, and its exchange produces quarks with large momenta
and therefore high energy.  In light-front coordinates the exchange of
a wee gluon changes the momentum of quarks by a small amount, allowing
them to have low energy.  As a result we find second-order two-body
interactions between low energy quarks generated by the removal of
coupling to high energy gluons, and these interactions are crucial for
producing a constituent picture in light-front coordinates.  Since
these interactions are dominated by the exchange of wee gluons, I will
concentrate on the part of the quark-antiquark interaction that
diverges as the longitudinal momentum exchange goes to zero.

In the case of gluon exchange between a quark and antiquark, we need
the coupling coherent solution in Eq. (7).  The exchange of high
energy gluons is removed by the cutoff and replaced in second order by
a direct interaction with cutoffs projecting on all intermediate state
energies above the cutoff.  We must add the canonical instantaneous
gluon exchange interaction to this induced interaction.  The induced
interaction is

\begin{eqnarray}
{V}_{\Lambda} &=&
 - 4 g_{\Lambda}^2 C_F \sqrt{p_1^+ p_2^+ k_1^+ k_2^+}\;
{q_\perp^2 \over (q^+)^3}
\nonumber \\ &&~~\times
\Biggl[ {\theta\bigl(|p_1^- -p_2^- -q^-| -\Lambda^2 / {\cal P}^+ \bigr)
\;\; \theta\bigl(|p_1^- -p_2^- -q^-|- |k_2^- -k_1^- -q^-| \bigr)
\over p_1^- -p_2^- -q^-} \nonumber \\ &&~~~~
+{\theta\bigl(|k_2^- -k_1^- -q^-| - \Lambda^2 / {\cal P}^+ \bigr)
\;\; \theta\bigl( |k_2^- -k_1^- -q^-| - |p_1^- -p_2^- -q^-| \bigr)
\over k_2^- -k_1^- -q^-} \Biggr]
\nonumber \\
&&~~~~~~~~~~~~\times
\theta\Bigl(\Lambda^2 / {\cal P}^+-\mid p_1^-+k_1^--p_2^--k_2^-
\mid \Bigr) \;.
\end{eqnarray}

\noindent Here the initial and final quark (antiquark) momenta are
$p_1$ and $p_2$ ($k_1$ and $k_2$), and the exchanged gluon momentum is
$q$.  The energies are all determined by the momenta,
$p_1^-=p^2_{\perp 1}/p_1^+$, etc.  This part of the interaction is
independent of the spins, which remain unchanged.

This interaction can be further simplified for our analysis by noting
that we are interested only in its most singular part, for which $q^+$
is extremely small.  In this case $|q^-|$ is much larger than $p_i^-$
and $k_i^-$, leading to the approximation

\begin{eqnarray}
{V}_{\Lambda} &\approx&
4 g_{\Lambda}^2 C_F \sqrt{p_1^+ p_2^+ k_1^+ k_2^+}\;
\Biggl({1 \over q^+}\Biggr)^2\;
\theta\bigl(|q^-| -\Lambda^2 / {\cal P}^+ \bigr)
\nonumber \\
&&~~~~~~~~~~~~\times
\theta\Bigl(\Lambda^2 / {\cal P}^+-\mid p_1^-+k_1^--p_2^--k_2^-
\mid \Bigr) \;.
\end{eqnarray}

\noindent The entire analysis can be made without making this
approximation, and the results are the same.

We also need the instantaneous gluon exchange interaction,

\begin{eqnarray}
V_{instant} &=& - 4 g_{\Lambda}^2 C_F \sqrt{p_1^+ p_2^+ k_1^+ k_2^+}
\;\Biggl({1 \over q^+}\Biggr)^2  \nonumber \\
&&~~~~~~\times
\theta\Bigl(\Lambda^2 / {\cal P}^+ -\mid p_1^-+k_1^--p_2^--k_2^- \mid
\Bigr)
\;.
\end{eqnarray}

\noindent The final cutoff on each of these interactions is the same,
requiring the quark-antiquark energy to change by less than the
cutoff.  Since this final cutoff appears everywhere and is unimportant
for the discussion, I will drop it.  In addition to the cutoffs I have
displayed, the same cutoff on longitudinal momenta used in the last
Section must be added; so that all longitudinal momenta are required
to exceed $\epsilon {\cal P}^+$.

Adding the above interactions and inserting the cutoff on longitudinal
momenta we find

\begin{eqnarray}
V_{singular} &=&
 - 4 g_{\Lambda}^2 C_F \sqrt{p_1^+ p_2^+ k_1^+ k_2^+}\;
\Biggl({1 \over q^+}\Biggr)^2\;
\theta\bigl(\Lambda^2 / {\cal P}^+ - |q^-| \bigr) \;
\theta\bigl(|q^+|-\epsilon {\cal P}^+\bigr) \;.
\end{eqnarray}

\noindent The most singular part of the one-gluon exchange operator
cancels the instantaneous interaction above the cutoff, leaving us
with the instantaneous exchange potential below the cutoff.  If
$\Lambda \approx \Lambda_{QCD}$, we expect further gluon exchange to
be suppressed, and we are left with this singular interaction between
the quark and antiquark.

The next step in the analysis is to take the expectation value of this
interaction between arbitrary quark-antiquark states.  The first
cutoff forces $|q^+|/{\cal P}^+ > q_\perp^2/\Lambda^2$, and the second cutoff
forces $|q^+|> \epsilon {\cal P}^+$.  We see that $|q^+|$ can reach
its lower limit only when $q_\perp^2 < \epsilon \Lambda^2$, and as
$\epsilon \rightarrow 0$ the singularity is suppressed because of this
phase space restriction; but it is not removed.  The expectation value
is

\begin{eqnarray}
\langle \Psi_2(P)|V_{singular}|\Psi_1(P)\rangle &=&
-4 g_\Lambda^2 C_F \int {dp^+_1 d^2p_{\perp 1} \over 16\pi^3}
{dp^+_2 d^2p_{\perp 2} \over 16\pi^3} \phi_2^*(p_2) \; \phi_1(p_1)
\nonumber \\
&&~~~~~~\times~\Biggl({1 \over q^+}\Biggr)^2\;
\theta\bigl(\Lambda^2 / {\cal P}^+ - |q^-| \bigr) \;
\theta\bigl(|q^+|-\epsilon {\cal P}^+\bigr) \;,
\end{eqnarray}

\noindent where as usual $q=p_1-p_2$ and $q^-=q_\perp^2/q^+$.  The
wave functions for the relative motion of the quark-antiquark pair are
$\phi_1$ and $\phi_2$, and I have suppressed their dependence on the
total momentum $P=p_1+k_1$.  I have not displayed the delta function
normalization associated with center-of-mass motion.  To evaluate the
singular part of this integral, change variables to

\begin{equation}
Q={p_1+p_2 \over 2} \;\;,\;\; q=p_1-p_2 \;,
\end{equation}

\noindent and expand the wave functions about $q=0$.  Only the leading
term diverges, and it is

\begin{eqnarray}
\langle \Psi_2(P)|V_{singular}|\Psi_1(P)\rangle &=&
-4 g_\Lambda^2 C_F \int {dQ^+ d^2Q_{\perp} \over 16\pi^3}
\phi_2^*(Q) \; \phi_1(Q) \nonumber \\
&&~~~~\times~ \int {dq^+ d^2q_{\perp} \over 16\pi^3}\;
\Biggl({1 \over q^+}\Biggr)^2\;
\theta\bigl(\Lambda^2 / {\cal P}^+ - |q^-| \bigr) \;
\nonumber \\ &&~~~~~~~~\times~
\theta\bigl(|q^+|-\epsilon {\cal P}^+\bigr) \;
\theta\bigl(\eta {\cal P}^+-|q^+| \bigr) \; \;+\;finite\;.
\end{eqnarray}

\noindent $\eta$ is an arbitrary constant that simply prevents $|q^+|$
from becoming too large, and it does not matter since the divergence
comes only from small $|q^+|$.  Completing the final integral we obtain

\begin{eqnarray}
\langle \Psi_2(P)|V_{singular}|\Psi_1(P)\rangle &=&
-{g_\Lambda^2 C_F \Lambda^2 \over 2 \pi^2 {\cal P}^+}
\log\Bigl( {1 \over
\epsilon} \Bigr) \int {dQ^+ d^2Q_{\perp} \over 16\pi^3}
\phi_2^*(Q) \phi_1(Q)  +finite.
\end{eqnarray}

Unless $\phi_1$ and $\phi_2$ are the same, this vanishes by
orthogonality.  If they are the same, this is exactly the same
expression we obtain for the expectation value of the quark plus
antiquark divergent mass operators; except with the opposite sign.
Therefore, there is a divergence in the quark-antiquark interaction
that is independent of their relative motion and which exactly cancels
the divergent masses!  These cancellations only occur for color
singlets, and they occur for any color singlet state with an arbitrary
number of quarks and gluons.  Moreover, these cancellations appear
directly in the hamiltonian matrix elements, so we can take the
$\epsilon \rightarrow 0$ limit before diagonalizing the matrix.

This is half of the simple confinement mechanism.  At this point it is
possible to obtain finite mass hadrons even though the parton masses
diverge.  However, since the cancellations are independent of the
relative parton motion, we must study the residual interactions to see
if they are confining.  Since I am interested in the long-range
interaction, I will study the fourier transform of the potential and
compute $V(r)-V(0)$ so that the divergent constant in which we are no
longer interested is canceled.  The details of computing a fourier
transform are not illuminating, so I will simply list the results.

\begin{equation}
V_{singular}(r)-V_{singular}(0) \rightarrow
{g_\Lambda^2 C_F \Lambda^2 \over 4
\pi^2 {\cal P}^+} \; \log\bigl(|x^-|\bigr) \;,
\end{equation}

\noindent when $x_\perp=0$ and $|x^-| \rightarrow \infty$; and

\begin{equation}
V_{singular}(r)-V_{singular}(0) \rightarrow
{g_\Lambda^2 C_F \Lambda^2 \over 2
\pi^2 {\cal P}^+} \; \log\bigl(|x_\perp|\bigr) \;,
\end{equation}

\noindent when $|x_\perp| \rightarrow \infty$ and $x^-=0$.  This
potential is not rotationally symmetric,  but it diverges
logarithmically in all directions.

If the potential is not rotationally symmetric, how can rotational
symmetry be restored?   To answer this question, remember that the
generators of rotations in light-front field theory contain
interactions that change parton number.  We expect the physical states
in which a quark and antiquark are separated by a large distance to
contain gluons.  There is no reason to assume that the gluon content
of these states is the same when the state is rotated, so rotational
symmetry will be restored in highly excited states only if we allow
additional partons.  This complicates our attempt to derive a
constituent picture, but we only need the constituent picture to work
well for low-lying states.  The intermediate range part of the
potential is rotationally symmetric, and we may expect the ground
state hadrons to be dominated by the valence configuration.

Isn't the confining potential supposed to be linear and not
logarithmic?  As far as I know there is no conclusive evidence that
the long-range potential is linear, and heavy quark phenomenology
shows that a logarithmic potential can work quite well. [12] The fact
is that we know extremely little about highly excited states with
large color dipole moments, and it is not clear that measurable
quantities are ever sensitive to such states.  In any case, I do not
want to argue that these calculations show that the long-range
potential in light-front QCD is logarithmic.  Higher order corrections
could produce powers of logarithms that add up to produce a linear
potential.

The important point with which I will conclude is that $H$ contains a
confining interaction that we are free to include in $H_0$, giving us
some hope of finding a reasonable bound state perturbation theory for
hadrons that resembles the bound state perturbation theory that has
been successfully applied to the study of atoms.

\vskip.3in
\noindent {\bf Acknowledgment}
\vskip.1in

I would like to thank the organizers, and especially Stan G{\l}azek,
for hosting an excellent conference.  I had not anticipated the
opportunity of climbing in the beautiful Tatra mountains under a full
moon.  I have profited from conversations with many people,
particularly Ken Wilson, Stan G{\l}azek and Martina Brisudova; and I
gained new insights from conversations with Leonard Susskind and
others at the conference.  This work was supported by the National
Science Foundation under grant PHY-9409042.

\vskip.3in
\noindent {\bf References}
\vskip.1in

\noindent
1. K. G. Wilson, T. S. Walhout, A. Harindranath, W.-M. Zhang, R. J.
Perry, St. D. G{\l}azek, Phys. Rev. {\bf D49}, 6720 (1994).

\noindent
2. K. G. Wilson and J. B. Kogut, Phys. Rep. {\bf 12}, 75 (1974).

\noindent
3. K. G. Wilson, Rev. Mod. Phys. {\bf 47}, 773 (1975).

\noindent
4. R. J. Perry, Ann. Phys. {\bf 232}, 116 (1994).

\noindent
5. St. D. G{\l}azek and K.G. Wilson, Phys. Rev. {\bf D48}, 5863
(1993); Phys. Rev. {\bf D49}, 4214 (1994).

\noindent
6. R. J. Perry and K. G. Wilson, Nucl. Phys. {\bf B403}, 587 (1993).

\noindent
7. R. J. Perry, {\it Hamiltonian Light-Front Field Theory and Quantum
Chromodynamics}, to appear in the proceedings of {\it Hadrons 94},
Gramado, Brasil, April, 1994.

\noindent
8. K. G. Wilson and D. G. Robertson, {\it Light-Front QCD and the
Constituent Quark Model}, these proceedings.

\noindent
9. S. J. Brodsky and G. P. Lepage, in {\it Perturbative quantum
chromodynamics}, A. H. Mueller, Ed.  (World Scientific, Singapore,
1989).

\noindent
10. W. M. Zhang and A. Harindranath, Phys. Rev. {\bf D48}, 4868; 4881;
4903 (1993).

\noindent
11. R. J. Perry, Phys. Lett. {\bf 300B}, 8 (1993).

\noindent
12. C. Quigg and J.L. Rosner, Phys. Lett. {\bf 71B}, 153 (1977).

\end{document}